\documentclass[preprint, 5p,twocolumn]{elsarticle}

\usepackage[utf8]{inputenc}
\usepackage[english]{babel}
\usepackage{svg}
\usepackage{graphicx} % Include figure files
\usepackage{bm} % bold math
\usepackage{siunitx}
\usepackage{blindtext}
\usepackage{mathtools}
\usepackage{amsmath}
\usepackage{amsthm}
\usepackage{amssymb}
\usepackage{hyperref}
\usepackage{soul}
\usepackage{xcolor}
\usepackage{booktabs}
\usepackage[para,flushleft]{threeparttable}
\usepackage{multirow}

\usepackage[caption=false]{subfig}

\newcommand{\R}{\mathbb{R}}
\newcommand{\N}{\mathbb{N}}

\newcommand{\Rmeas}{\mathcal{R}}
\newcommand{\insitu}{\textit{in situ} }
\newcommand{\Insitu}{\textit{In situ} }
\newcommand{\ie}{i.e.\ }
\newcommand{\eg}{e.g.\ }
\newcommand{\viz}{viz.\ }
\renewcommand{\Re}{\operatorname{Re}}

\newcommand{\SIinterval}[3]{{[#1, #2]\,\si{#3}}}

\begin{document}
\begin{frontmatter}
	%\title{Phase Reconstruction of a Cu(001) Seed Layer with Stepwise Epitaxially Grown Fe as Reference Layers from \insitu Polarized Neutron Reflectometry Data}
	
\title{Phase Reconstruction of a Cu(001) Seed Layer from \insitu Polarized Neutron Reflectometry Data using Fe Reference Layers}
\author[TUM]{Alexander Book}
\ead{alexander.book@frm2.tum.de}
	
\author[TUM,PSI_alt]{Sina Mayr}
\fntext[PSI_alt]{Present address: Paul Scherrer Institut, 5232 Villigen PSI, Switzerland}
	
\author[PSI]{Jochen Stahn}
	
\author[TUM]{Peter Böni}
	
\author[TUM,CAS,SNSSC]{Wolfgang Kreuzpaintner}
	
%\address[MLZ]{Forschungs-Neutronenquelle Heinz Maier-Leibnitz (FRM II), Technische Universität München, Lichtenberg Strasse 1, 85748 Garching, Germany}
\address[TUM]{Technische Universität München, Physik-Department E21, James-Franck-Str. 1, 85748 Garching, Germany}
\address[PSI]{Laboratory for Neutron Scattering and Imaging, Paul Scherrer Institut, 5232 Villigen PSI, Switzerland}
\address[CAS]{Institute of High Energy Physics, Chinese Academy of Sciences (CAS), Beijing 100049, China}
\address[SNSSC]{Spallation Neutron Source Science Center (SNSSC), Dongguan 523803, China}
	
\date{\today}

\begin{abstract}
	The reconstruction of the complex reflection coefficient obtained by \insitu Polarized Neutron Reflectometry is presented. Using the reference layer method, with a magnetic Fe layer, the phase information of the underlying Cu(001) seed layer sample structure is successfully retrieved and its scattering length density is calculated and compared with results obtained from traditional fitting. Two different reference layer approaches for retrieving the phase information are compared.% and the effect of resolution and Poisson noise on the reconstructed reflection is discussed.
\end{abstract}
	

\end{frontmatter}
%\maketitle

\section{Introduction}
Polarized Neutron Reflectometry (PNR) is an indispensable scattering technique to probe structural and magnetic properties of thin films and heterostructures. The desired physical parameters are commonly extracted by fitting an appropriate theoretical model to the measured reflectivity. However, without any \textit{a-priori} knowledge of the sample structure, a manifold of scattering potentials can yield the same reflectivity since the phase information is not preserved in the measurement. This ambiguity is known as the phase problem.

Various solutions to the phase problem in neutron and x-ray reflectometry have been proposed: \citeauthor{fiedeldey_proposal_1992} \cite{fiedeldey_proposal_1992} use an ordinary differential equation involving the dwell time (\viz the time the neutron ``stays'' inside the sample) to solve for the reflection phase. \citeauthor{clinton_phase_1993} \cite{clinton_phase_1993} uses a logarithmic dispersion relation to associate the phase with the reflectivity in the Born-Approximation. Experimentally oriented solutions usually vary known segments of the sample to constrain the phase. 

Another idea consists of utilizing specific materials which exhibit resonance effects for wavelengths close to absorption edges, mainly for x-rays \cite{sanyal_fourier_1993}. These materials serve as references which vary their SLD depending on the wavelength while the SLD of the unknown part of the sample is constant. For neutron reflectometry in the thermal to epithermal range, rare earths are suitable candidates as absorbing materials \cite{lynn_resonance_1990} and corresponding experiments have been performed \cite{salamatov_use_2016, nikova_experimental_2019, nikova_development_2020}.

In the reference layer method \cite{majkrzak_exact_1995, de_haan_retrieval_1995} one varies the scattering length density (SLD) by changing the magnetization of the sample \cite{kirby_phase-sensitive_2012}, by using buried or fronting layers of different SLDs \cite{majkrzak_phase_1998, majkrzak_progress_2009} or by a variation of sample's surroundings \cite{majkrzak_experimental_2000}. Other approaches use an adapted version of the reference layer method by measuring the polarization of the reflected neutrons instead of the reflectivity for assessing the phase below the critical edge \cite{leeb_determination_1998, kasper_phase_1998}.

To uniquely determine the scattering potential for \insitu PNR, we reconstructed the phase information of a model system, \ie a Cu seed layer, by applying the fronting reference layer method \cite{majkrzak_phase-sensitive_2003}. We successively increased the thickness of the top Fe reference layer and after each Fe deposition step we inspected the sample by \insitu PNR. The usage of Fe as a fronting material allows us to increase the contrast by magnetizing the reference layer, thus improving the quality of the phase information. We present two reference layer approaches, \viz the remnant Fe layer approach in Section \ref{sec:remnant_layer_up} and an improvement by incorporating the remnant Fe layer into the reference in Section \ref{sec:combination}. Finally, we extrapolated the reflection to further assess the results from the inverted SLD.

This work demonstrates a proof of concept of the application of a phase reconstruction technique for \insitu PNR. It is shown that \insitu PNR is well suited for phase reconstruction since a large number of reference layers can be created and measured with the very same underlying sample.

\section{Experimental Procedure} 
\begin{figure}
	\centering
	\includegraphics[width=\columnwidth]{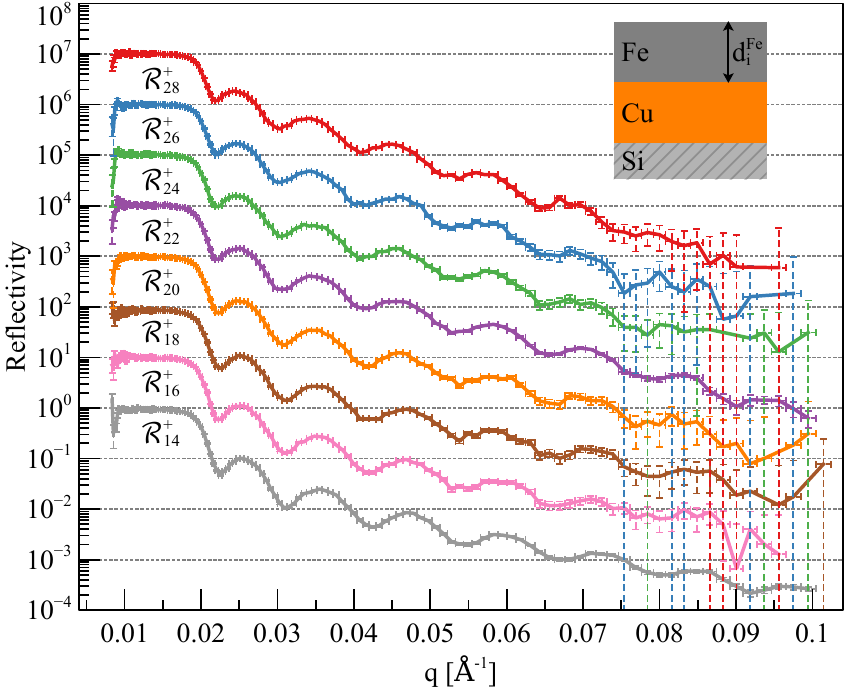}
	\caption{
		Reflectivity data of the Si/Cu/Fe sample with varying Fe layer thickness. Each reflectivity is shifted by one order of magnitude for clarity. The reflectivity curve $\Rmeas_n^+$ corresponds to the sample with $n$ monolayers of Fe on top with the Fe layer being magnetized parallel to the neutron spin polarization. The inset depicts the sample's layer structure.
	}
	\label{fig:reflectivity_data}
\end{figure}

A $\SI{45}{nm}$ thick Cu seed layer was grown on a $\SI{2x2}{cm}$ Si(001) substrate using DC magnetron sputtering in an ultrahigh-vacuum deposition chamber \cite{schmehl_design_2018, ye_design_2020}. Subsequently, a $\SI{7}{nm}$ thick Fe thin film was added on the Cu layer in 28 deposition steps with a growth rate of approximately one monolayer of Fe per deposition step. PNR measurements are carried out after each deposition step; after the 14th deposition, only every second step is monitored by PNR \cite{kreuzpaintner_situ_2017}. The deposition and PNR measurements were performed on the neutron reflectometer Amor at the Swiss Spallation Neutron Source (SINQ), Paul Scherrer Institut, Villigen \cite{stahn_focusing_2016}. The resulting spin-up (+) and spin-down (-) reflectivity is denoted by $\Rmeas_i^\pm$. Figure \ref{fig:reflectivity_data} depicts the performed PNR measurements for the deposition steps $14$ to $28$ which will will be used for reconstructing the reflection of the Cu seed layer. The reflectivity $\Rmeas_{14}$ is shown for completeness.

To reconstruct the Cu seed layer, we employ the reference layer method described in \cite{majkrzak_phase-sensitive_2003} by using the top Fe layer as the reference. In the process of reconstruction, the contribution of the Fe layer on the reflection will be eliminated and only the reflection of Si/Cu will be retrieved. For the reconstruction we use seven different sample configurations, yielding in total 14 PNR measurements. 

It is noted that the requirements \cite{sacks_reconstruction_1993, majkrzak_phase-sensitive_2003} of the SLD for the reconstruction are all fulfilled, namely: All the constituents of the sample (Si, Cu, Fe) have a SLD with non-negative real part (hence no bound states \cite{sacks_reconstruction_1993}) and the imaginary part is sufficiently small ($\approx 4$ orders of magnitude smaller compared to the real part) which justifies the assumption that the SLD is completely real valued. The sample is also of finite extent regarding its top surface and we assume that the unknown portion of the sample (\ie Si/Cu) remains constant in each Fe deposition step.

\section{Data Analysis}
The reflection $R$ is the complex amplitude of the wave function of the neutrons $\psi(z) \sim 1e^{-iqz} + R e^{iqz}$ as $z\to \infty$, hence the neutrons are impinging on the scattering potential from the right in Figure \ref{fig:schematic_references}. The reflectivity $|R|^2$ is the squared modulus of the reflection $R$ and the phase $\phi = \arg{(R)}$ is correlated with the reflection by $R = |R| e^{i\phi}$.

In general it is sufficient to use three PNR measurements of the same sample with different reference layers to unambiguously reconstruct the reflection $R$ from inversion of a $3\times3$ ``constraint''-matrix and to obtain a reflection parameter vector $\Theta = (\alpha, \beta, \gamma) \in \R^3$. Here, $\alpha, \beta \text{ and } \gamma$ originate from the elements of the transfer-matrix \cite{majkrzak_exact_1995}. In the matrix formulation of the one dimensional Schrödinger equation the transfer-matrix relates the reflected to the transmitted plane wave within a thin film. The reflection $R$ is calculated from the reflection parameter vector $\Theta$ by
\begin{equation}
R = \frac{\beta - \alpha - 2i\gamma}{\alpha + \beta + 2}.
\end{equation}

We can incorporate $N\in\N$ measurements by using a chi-squared ($\chi^2$) ansatz 
\begin{equation}\label{eq:chi_squared}
\chi^2(\Theta) = \sum_{i=1}^{N} \left(\frac{L_i \cdot \Theta - b(|R_i|^2)}{\sigma_{b(|R_i|^2)}}\right)^2,
\end{equation}
and solve for the $\chi^2$-minimizing reflection parameter $\Theta$. The minimizer of $\chi^2$ can be analytically computed by the linear least squares method. The solution exists and is unique since there are more measurements than unknowns and the least squares matrix $L = (L_i)_i^N$ has full rank. In this ansatz, the vector $L_i \in \R^3$ contains the reference layer information (\ie the complete SLD profile of the reference layer) and $b(|R_i|^2)$ is a transformed reflectivity measurement with standard deviation $\sigma_{b(|R_i|^2)}$: 
\begin{equation}\label{eq:reflectivity_transformation}
b\left(|R_i|^2\right) \sim \frac{1+|R_i|^2}{1-|R_i|^2}
\end{equation}
This shows why the reconstruction can only work outside the total reflection regime, since the denominator $1 - |R_i|^2 = 0$ for total reflection. 

With the usage of the $\chi^2$ ansatz, the uncertainty in the reflectivity data can be taken into account and the resulting reflection can be reconstructed more precisely, due to better statistics of the multiple reflectivity data sets and due to the reduction of the error in the SLD profile of the reference layer. In particular the last point is worth noting: It is highly complicated to precisely determine the shape of the reference layers which are required for the reconstruction. However, by incorporating multiple measurements, the importance of a precise knowledge of the reference layer appears to decrease. This is a key property from which we take advantage in this paper.

The reflection is inverted using the Gelfand-Levitan-Marchenko (GLM) method \cite{chadan_inverse_1989} to retrieve the SLD. We use only the real part of the reflection, as applying the imaginary part resulted in singularities that cannot be resolved. It is noted that we can freely choose either one of the real or imaginary quantities, as they yield the same SLD \cite{sacks_reconstruction_1993}. The software used for the reconstruction and potential inversion is available freely \cite{direfl, dinv}.

\section{Results}\label{sec:results}

\begin{figure}
	\centering
	\subfloat{\includegraphics[scale=0]{fig_99_invisible.png}\label{fig:schematic_references:a}}
	\subfloat{\includegraphics[scale=0]{fig_99_invisible.png}\label{fig:schematic_references:b}}
	\includegraphics[width=1\columnwidth]{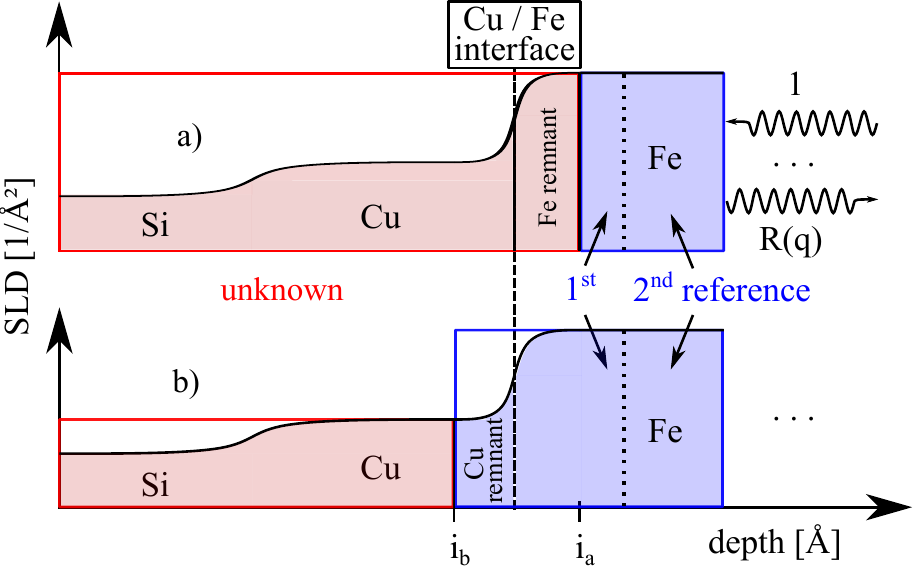}
	\caption{
		Schematic comparison of two different reference layer selections. The blue shaded area on the right depicts the reference layer whereas the left area is the unknown sample. \protect\subref*{fig:schematic_references:a}) The remnant Fe layer is shifted to the unknown part of the sample. \protect\subref*{fig:schematic_references:b}) The full Fe layer, containing interdiffusion and/or roughness of the Cu/Fe interface, is included into the reference.
	}
	\label{fig:schematic_references}
\end{figure}

The sample is virtually split into two parts at a depth $i$; an unknown part and a reference layer. The splitting point $i$ can be chosen arbitrarily, but only a few specific ones are physically meaningful (see Figure \ref{fig:schematic_references}). The variables $i_a$ and $i_b$ denote the depths inside the sample at which the SLD is completely characterized by only one of the contributions of Fe or Cu, respectively: 

Splitting the sample at $i = i_a$ assigns the remnant Fe layer to the unknown sample. This case is investigated in Section \ref{sec:remnant_layer_up}. The advantage of this approach is that it does not require knowledge of the Cu/Fe interface, however, due to the magnetism in the Fe layer, only data of one spin direction can be used. Alternatively, the splitting at $i = i_b$ allows both spin direction data to be used, since the unknown sample is non-magnetic, but this requires additional \textit{a-priori} knowledge of the Cu/Fe interface. 

In the following it is shown that using only measurements of one spin direction (Section \ref{sec:remnant_layer_up}) is sufficient to reconstruct the reflection. However, significant improvements in the quality of the reconstruction can be obtained by using the experimental data of both spin directions (Section \ref{sec:combination}).

It is noted that by specular reflectivity alone roughness and interdiffusion cannot be distinguished; henceforth the term roughness also includes interdiffusion and vice versa.

\subsection{Reconstruction Using a Remnant Fe Layer}\label{sec:remnant_layer_up}

The top Fe layer represents the reference, however the Fe layer thickness is constrained and a part of it (remnant Fe) is declared to belong to the unknown part (Figure \ref{fig:schematic_references:a}). It is noted that the remnant layer has a non-zero magnetization which prohibits the use of both spin state reflectivity measurements for reconstructing the reflection.

\begin{figure}
	\subfloat{\includegraphics[scale=0]{fig_99_invisible.png}\label{fig:reference_layers:a}}
	\subfloat{\includegraphics[scale=0]{fig_99_invisible.png}\label{fig:reference_layers:b}}
	\includegraphics[width=\columnwidth]{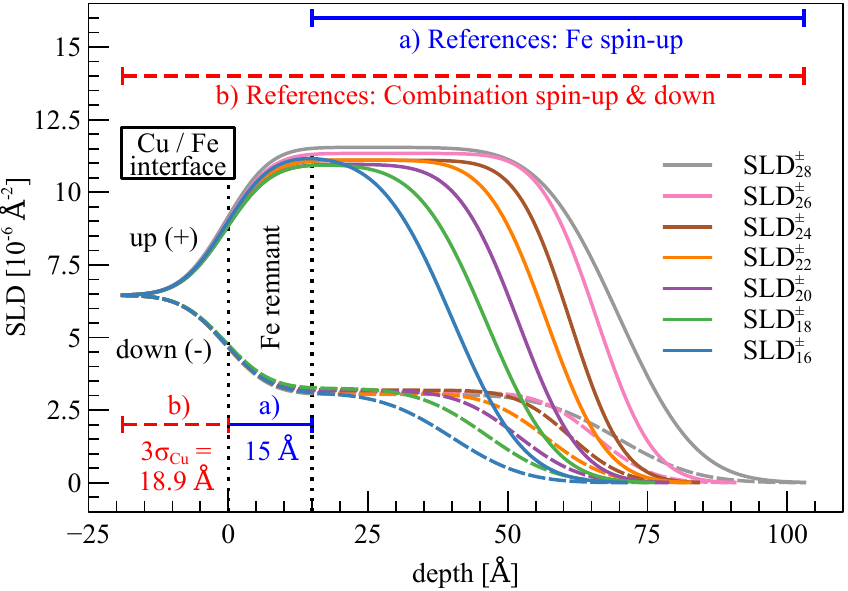}
	\caption{
		The SLD profiles of the reference layers, scaled such that a depth of $\SI{0}{\angstrom}$ refers to the Cu/Fe interface. \protect\subref*{fig:reference_layers:a}) Spin-up SLDs of reference layers with a remnant Fe layer (Section \ref{sec:remnant_layer_up}). \protect\subref*{fig:reference_layers:b}) Reference layers comprising the full Fe layer and including the Cu/Fe interface for both spin channels (Section \ref{sec:combination}).
	}
	\label{fig:reference_layers}
\end{figure}

\begin{table}
	\centering
	\caption{
		Fitted parameters of the Fe reference layer and of the unknown layer (from \cite{kreuzpaintner_situ_2017}). 
	}	
	\label{table:reference_layer_parameters}
		\begin{tabular}{l S[table-format=3.1] S[table-format=1.2] S[table-format=2.1] S[table-format=1.3]}
			\toprule
			Name & {thickness} & {density} & {roughness} & {magnetization}\\
			& {$d$ [\SI{}{\angstrom}]} & {$\rho$ [\SI{}{g/cm^3}]}  & {$\sigma$ [\SI{}{\angstrom}]} & {$M$ [\SI{}{\mu_B/atom}]}\\
			\midrule
			$\Rmeas_{28}$ & 70.3 & 7.16 & 11.0 & 2.03 \\
			$\Rmeas_{26}$ & 66.4 & 7.09 &  8.0 & 1.98 \\
			$\Rmeas_{24}$ & 61.7 & 7.01 &  7.5 & 1.93 \\
			$\Rmeas_{22}$ & 57.5 & 6.96 &  8.4 & 1.97 \\
			$\Rmeas_{20}$ & 52.5 & 6.94 &  8.7 & 1.92 \\
			$\Rmeas_{18}$ & 46.8 & 6.97 &  9.7 & 1.89 \\
			$\Rmeas_{16}$ & 40.6 & 7.05 & 10.1 & 1.98 \\
			$\Rmeas_{14}$ & 36.7 & 6.95 &  9.3 & 1.91 \\
			\toprule
			&  {$d$ [\SI{}{\angstrom}]} & {$\rho$ [\SI{}{g/cm^3}]} & {$\sigma$ $[\SI{}{\angstrom}]$} & {SLD $[\SI{}{\angstrom^{-2}}]$} \\
			\midrule
			Cu		& {451} 		& 8.82 	& 6.3 	& \SI{6.45e-6}{} \\
			Si		& $\infty$  & 2.33 	& 10.7 	& \SI{2.08e-6}{} \\
			\bottomrule
		\end{tabular}
\end{table}

Figure \ref{fig:reference_layers} depicts the SLD profile of the reference layers. Each reflectivity measurement $\Rmeas_i^\pm$ has a corresponding reference layer $\text{SLD}_i^\pm$ which is used for the reconstruction of the reflection. 
\Insitu thin film growth allows the thickness of the deposited Fe layer to be known \textit{a-priori}, however, the parameters used for describing the SLD of the Fe reference layer were extracted using a traditional fitting method \cite{kreuzpaintner_situ_2017}. 
These are listed in Table \ref{table:reference_layer_parameters} for $\Rmeas_{14}$ - $\Rmeas_{28}$.

Here, we use the reference layers configuration shown in Figure \ref{fig:reference_layers:a}. The roughness at the Cu/Fe interface is assumed to be smaller than $3 \sigma = \SI{15}{\angstrom}$. This restricts the usable reflectivity measurements to those in the deposition steps 6 to 28, \viz $\Rmeas_{6}$ - $\Rmeas_{28}$. To further simplify the shape of the reference layers only those with a thickness significant above $\approx \SI{35}{\angstrom}$ are considered, \viz $\Rmeas_{16}$ - $\Rmeas_{28}$. We selected the spin-up reflectivity $\Rmeas^+$ since it has better statistics for higher $q$ wave vector transfers. The case of using only spin-down reflectivities is given in \ref{appendix:fe_down_statistics}. 

%Using the reflectivity measurements $\Rmeas_{16}$ - $\Rmeas_{28}$ (Figure \ref{fig:reflectivity_data}]) and defining the reference layers as depicted in Figure \ref{fig:reference_layers:a}, 

\begin{figure}
	\subfloat{\includegraphics[scale=0]{fig_99_invisible.png}\label{fig:fe_up_remnant_reconstructed_phase}}
	\subfloat{\includegraphics[scale=0]{fig_99_invisible.png}\label{fig:fe_up_remnant_inverted_SLD}}
	\centering
	\includegraphics[width=\columnwidth]{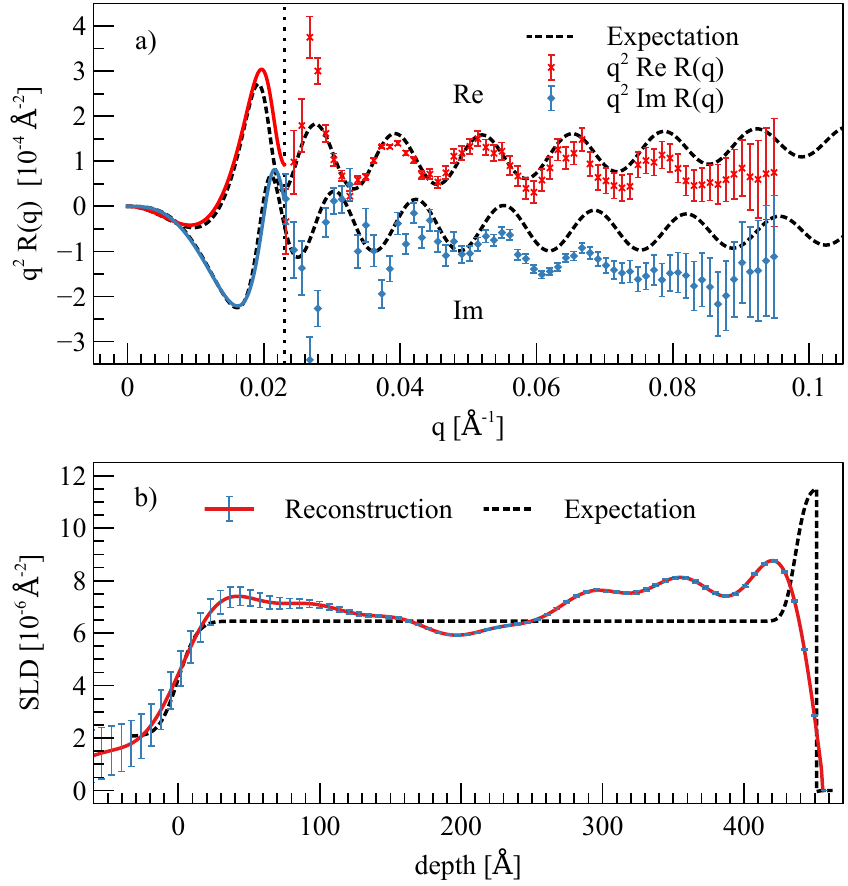}
	\caption{
		\protect\subref*{fig:fe_up_remnant_reconstructed_phase}) Reconstructed reflection multiplied with $q^2$ of the unknown Cu layer, including a remnant Fe layer (see figures \ref{fig:schematic_references:a} and \ref{fig:reference_layers:a}). The reflectivity data $\Rmeas_{16}^+$ - $\Rmeas_{28}^+$ is used for the reconstruction. The expected curve (dashed line) represents the reflection of the fitted model which is a $\SI{436}{\angstrom}$ thick Cu layer on a Si substrate with a $\SI{15}{\angstrom}$ thick magnetized Fe layer on top. The reflection below the critical edge $q \leq \SI{0.023}{\angstrom^{-1}}$ (dotted vertical line) is numerically retrieved and its error is set to zero.
		\protect\subref*{fig:fe_up_remnant_inverted_SLD}) Inverted SLD using the real part of the reflection in  \protect\subref{fig:fe_up_remnant_reconstructed_phase}.
	}
\end{figure}

Figure \ref{fig:fe_up_remnant_reconstructed_phase} shows the reconstructed reflection. 
%the reflection is reconstructed as shown in Figure \ref{fig:fe_up_remnant_reconstructed_phase}. 
It exhibits low frequency oscillations from the Cu layer and high frequency oscillations, which result from the measurement noise in the reflectivity (for more details see \ref{appendix:fe_down_statistics}). The expected reflection is the theoretically calculated reflection based on fitted parameters given in Table \ref{table:reference_layer_parameters}.

The reflection below the critical edge $q_c = \sqrt{16\pi\text{SLD}}$ was retrieved using the fixed-point algorithm described in \cite{book_retrieval_2020}. Note that this particular algorithm only guarantees a successful retrieval if the product of film thickness and minimal wave vector transfer is smaller than $2 \pi$. This is, however, not given in our case, as the Cu layer is thicker than $\SI{400}{\angstrom}$ and the minimal wave vector is $\approx \SI{0.023}{\angstrom^{-1}}$. To ensure the algorithm to converge, a $\SI{150}{\angstrom}$ thick portion of the unknown Cu layer was assigned to the SLD of bulk Cu ($\text{SLD}_{\mbox{Cu}} = \SI{6.554e-06}{\angstrom^{-2}}$ \cite{koester_neutron_1991}). It was positioned inside the Cu layer between $\SI{200}{\angstrom} \leq \mbox{depth} \leq \SI{350}{\angstrom}$. Variations of the position ($\pm \SI{50}{\angstrom}$) or thickness ($\SIrange{150}{250}{\angstrom}$) of the bulk Cu portion did not substantially change the retrieved reflection.

The corresponding inverted SLD is displayed in Figure \ref{fig:fe_up_remnant_inverted_SLD} and the dotted curve shows the slab model which is used for calculating the expected reflection. 
At the Si/Cu interface (between a depth of $\SI{-20}{\angstrom}$ and $\SI{20}{\angstrom}$) the inverted SLD matches the expected SLD remarkably well. 
However, the inverted SLD in general exceeds the expectation, except at the dip at the center $\approx \SI{200}{\angstrom}$ or at the remnant Fe layer $\approx \SI{450}{\angstrom}$. 
The quality of the reconstructed reflection is not sufficient to draw other conclusions beside estimating the Cu layer thickness.

\subsection{Reconstruction by Combination of Spin-Up and Down Measurements}\label{sec:combination}

To improve the quality of the reconstructed reflection, spin-down and spin-up measurements are used together. To increase the contrast and to allow for this approach, the remnant Fe layer needs to be part of the reference instead of the unknown sample. Only this allows the unknown part of the sample to be non-magnetic. Therefore, knowledge of the interface roughness of the Cu layer and density at the interface is needed, since these parameters determine the shape of the reference at the Cu/Fe interface. We use the fitted values for the interface (Table \ref{table:reference_layer_parameters}), however, it is noted that no substantial change in the reflection was observed if only approximate parameters ($\text{SLD}_{\text{Cu}} = \SI{6.554e-6}{\angstrom^{-2}}$, $\sigma_{\text{Cu/Fe}}=\SI{5}{\angstrom}$) are used instead. The reference layer SLDs as used in the following are shown in Figure \ref{fig:reference_layers:b}.

\begin{figure}
	\centering
	\subfloat{\includegraphics[scale=0]{fig_99_invisible.png}\label{fig:reconstructed_phase_combination}}
	\subfloat{\includegraphics[scale=0]{fig_99_invisible.png}\label{fig:reconstructed_SLD_combination}}
	\includegraphics[width=\columnwidth]{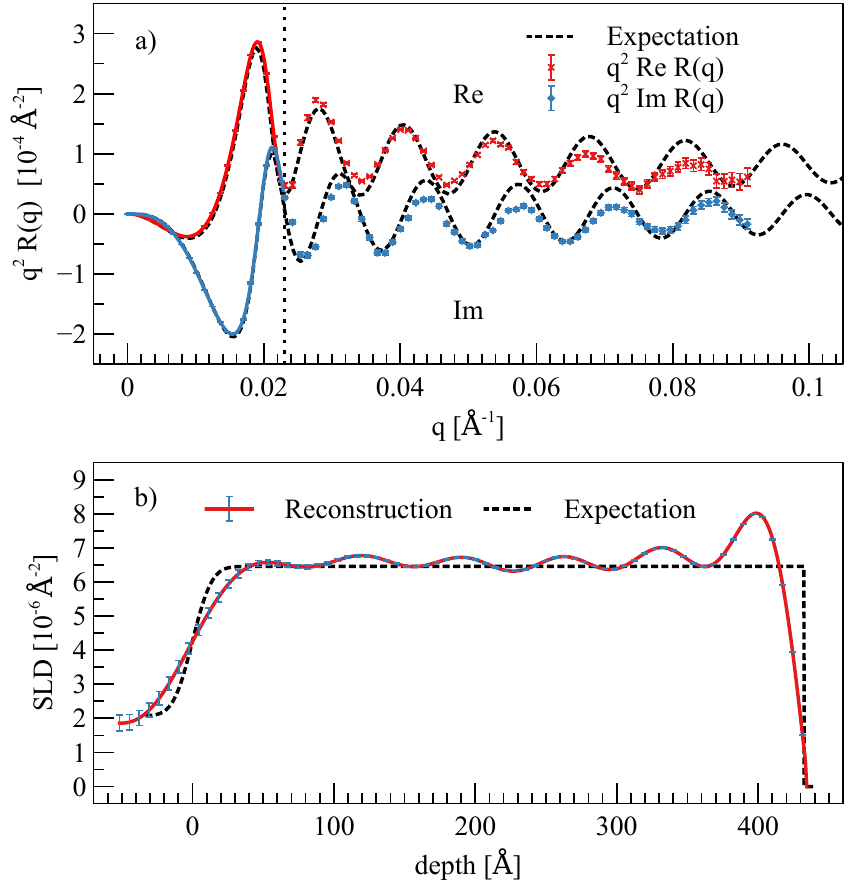}
	\caption{
		\protect\subref*{fig:reconstructed_phase_combination}) Reconstructed reflection multiplied with $q^2$ of the unknown Cu layer (see figures \ref{fig:schematic_references:b} and \ref{fig:reference_layers:b}), using the reflectivity data $\Rmeas_{16}^\pm$ - $\Rmeas_{28}^\pm$. The dashed curves represent the expected reflection of the fitted model. The reflection for $q \leq \SI{0.023}{\angstrom^{-1}}$ (dotted line) is numerically retrieved.
		\protect\subref*{fig:reconstructed_SLD_combination}) Inverted SLD using the reflection shown in \protect\subref{fig:reconstructed_phase_combination}. The dotted line represents the potential retrieved via classical fitting. 
	}
	%\label{fig:reconstructed_SLD_combination}
\end{figure}

The reconstructed reflection is given in Figure \ref{fig:reconstructed_phase_combination}. 
The missing reflection $R(q)$ for $q \in \SIinterval{0}{0.023}{\angstrom^{-1}}$ is retrieved by the fixed-point algorithm as described in Section \ref{sec:remnant_layer_up}. 
Figure \ref{fig:reconstructed_SLD_combination} shows the resulting inverted SLD. 
By using both spin direction measurements, the reconstructed SLD more closely resembles the fitted SLD, if compared to previous results (Figure \ref{fig:fe_up_remnant_inverted_SLD})), and the pronounced dip in the center of the potential at a depth of \SI{200}{\angstrom} is reduced.

\subsection{Fitting the Reflection}\label{sec:reflection_fit}
To eliminate the remaining oscillations in the inverted SLD, knowledge of the reflection for a larger $q$ range is required \cite{berk_statistical_2009}. The truncation of the reflection is currently one of the reasons why an even more precise potential inversion is not yet possible. Here, we demonstrate how an extrapolation of the given data allows the oscillations in the resulting potential to be eliminated: We fit a model function to the reflection data multiplied with $q^2$ and extrapolate the fitted model parameters to $q = \SI{5}{\angstrom^{-1}}$. As a model function for the real part of the reflection we use 
\begin{equation}
\label{eq:reflection_model}
\begin{split}
	q^2\Re R(q) &= O(q) - A(q) \cos \left(q d + \phi_0 + \phi_1 q^{-1} \right) ,\\
	O(q) &= 4\pi \rho + a_1q^{-1},\\
	A(q) &= \left(4\pi\rho_\Delta + \frac{a_0}{q - q_c}\right) e^{-q^2\sigma^2}.
\end{split}
\end{equation}
%with $C(q) = 4\pi \text{SLD}_{\text{Si}} + \frac{a_1}{q+q_c}$ is an offset and $A(q) = (c_0 + \frac{a_0}{q - q_c}) e^{-q^2\sigma^2}$ is a suitable representation of the amplitude.
The amplitude $A(q)$ and offset $O(q)$ function are selected heuristically such that they agree with a single layer model in the Born-Approximation when taking the limit $q\to\infty$.
The parameter $\sigma$ is the roughness or interdiffusion of the Cu/Si interface, $d$ determines the Cu layer thickness, $\phi_1 q^{-1}$ is a first order correction term to take the non-linearity of the periodicity into consideration. 
The offset $\rho$ is the SLD of Cu and $\rho_\Delta$ is the difference in the SLD of Cu and Si (see Table \ref{table:reference_layer_parameters}). 
The remaining parameters have no physical relevance. 
The best least-squares fitting parameters are listed in Table \ref{table:parameters_fit} in \ref{appendix:additional_table}.

\begin{figure}
	\centering
	\subfloat{\includegraphics[scale=0]{fig_99_invisible.png}\label{fig:scattering_potential_fitted_phase:a}}
	\subfloat{\includegraphics[scale=0]{fig_99_invisible.png}\label{fig:scattering_potential_fitted_phase:b}}
	\includegraphics[width=\columnwidth]{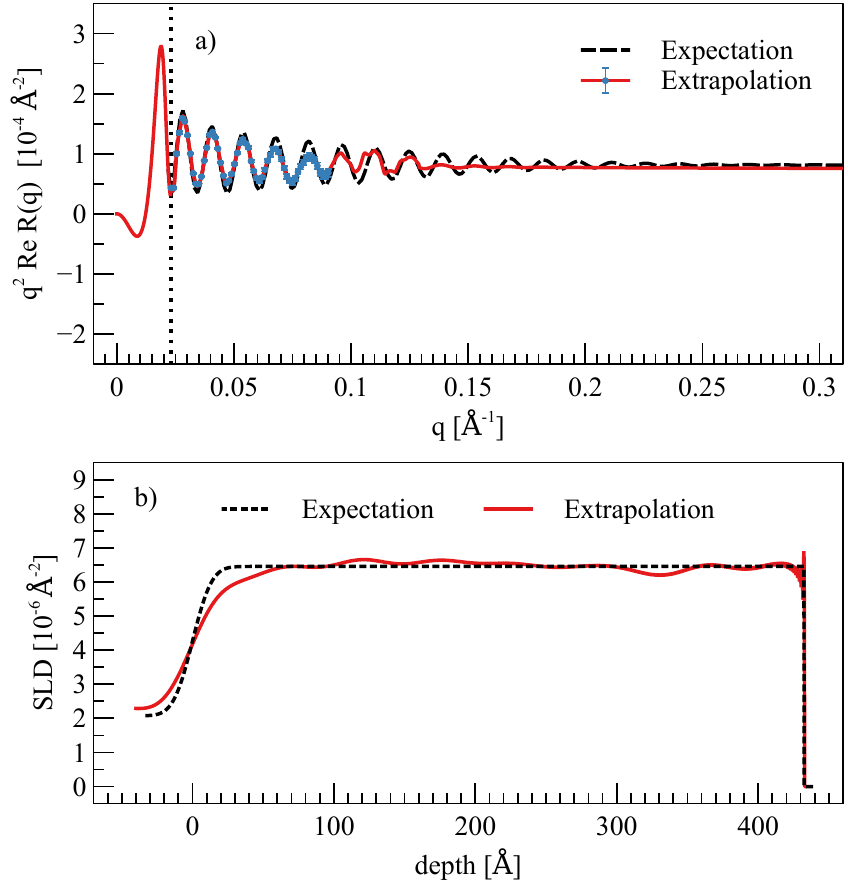}
	\caption{
		\protect\subref*{fig:scattering_potential_fitted_phase:a}) Extrapolated real part of the reflection multiplied with $q^2$ (solid curve) using model \eqref{eq:reflection_model}. The reflection below $q = \SI{0.023}{\angstrom^{-1}}$ (dotted line) is numerically retrieved and above $q = \SI{0.085}{\angstrom^{-1}}$ is extrapolated. The blue dots correspond to the reconstructed reflection from measurement data.
		\protect\subref*{fig:scattering_potential_fitted_phase:b}) Inverted SLD using the extrapolated reflection (see \protect\subref{fig:scattering_potential_fitted_phase:a}) $R(q)$ for $\SI{0}{\angstrom^{-1}} \leq q \leq \SI{5}{\angstrom^{-1}}$. 
	}
	\label{fig:scattering_potential_fitted_phase}
\end{figure}

The extrapolated reflection is depicted in Figure \ref{fig:scattering_potential_fitted_phase:a}. A clear deviation from the expected reflection can be observed, especially in the limit $q \to \infty$. Nevertheless, inverting the extrapolated reflection yields the SLD shown in Figure \ref{fig:scattering_potential_fitted_phase:b}. The small and sharp peak at the Cu/Fe interface at a depth of $\approx \SI{430}{\angstrom}$ is due to the Gibbs-phenomena \cite{hewitt_gibbs-wilbraham_1979}

\section{Discussion}
The reconstruction of an Fe remnant layer shows the possibility to use monolayers of Fe as a reference and to successfully retrieve the reflection. However, we observed high frequency oscillations over the entire $q$ range under investigation (Figure \ref{fig:fe_up_remnant_reconstructed_phase}) if compared with the expected reflection. We suspect the Poisson noise in the reflectivity data to cause the oscillations in the reconstructed reflection. This assumption is supported by the observation of a similar behavior when only simulated data is used. Simulations comparing the effect of the resolution onto the reconstructed reflection (\ref{appendix:resolution}) suggest that the resolution does not contribute to the high frequency oscillations, but that it only dampens the amplitude of the reflection. 

The large errors at high $q$ originate from the unsatisfactory signal-to-noise ratio of the reflectivity measurements when the reflectivity is of the same order of magnitude as the background. Applying the remnant layer approach to spin-down reflectivity measurements yields larger errors in the reflection (Figure \ref{fig:fe_down_remnant_reconstructed_phase} in \ref{appendix:fe_down_statistics}) than applying it to the spin-up reflectivity data. \ref{appendix:fe_down_statistics} discusses the effect of noise in more detail. Nonetheless, the retrieved profile shows in general a good agreement with the expected slab model, except at the jump discontinuity (Figure \ref{fig:fe_up_remnant_inverted_SLD}) at a depth of \SIrange{400}{450}{\angstrom}.

Improvements of the reconstruction can be achieved by combining spin-up and down reflectivity measurements, although the parameters of the interface between the unknown part of the sample to the references have to be estimated, which is in principle an information that one wants to obtain from the experiment. 

Nevertheless, the advantage of using the combined approach is the much higher contrast of the SLDs, which result from using magnetic materials and all the available PNR data. Hence, the reconstructed reflection has a strongly improved accuracy and agreement with the expected reflection, which yields a more accurate inverted SLD. It is noted that the errors in Figure \ref{fig:reconstructed_SLD_combination} are calculated with estimated errors in the low $q$ reflection regime (see \ref{appendix:error_estimation} for details), whereas the errors in Figure \ref{fig:fe_up_remnant_inverted_SLD} are neglected.

\begin{table}
	\centering
	\caption{
		Comparison of the layer parameters obtained by using different reference layers. 
	}
	\label{table:comparison_layer_parameters}
	\begin{threeparttable}
	\begin{tabular}{l S[table-format=2.1] S[table-format=2.1] S[table-format=1.2] S[table-format=3.0]}
		\toprule
		\multirow{2}{*}{Method}	&		\multicolumn{2}{c}{Si}	& \multicolumn{2}{c}{Cu}\\
					\cmidrule(lr){2-3} 		  \cmidrule(lr){4-5}
		&	{$\rho$ [$\SI{}{g/cm^3}$]} & {$\sigma$ [$\SI{}{\angstrom}$]} & {$\rho$ [$\SI{}{g/cm^3}$]}	&	{$d$ [$\SI{}{\angstrom}$]}\\
		\midrule
		Only spin-up	& 2.01	&	24.1	&	9.63	&	427	\\
		Spin-up \& down	& 2.20	&	17.7	&	8.93	&	426	\\	
		Extrapolation	& 2.57	&	12.3	&	8.83	&	432	\\
		\midrule
		Traditional fit	& 2.33	&	10.7	&	8.82	&	432\tnote{a} \\
		\bottomrule
	\end{tabular}
	\begin{tablenotes}\footnotesize
		\item[a] after subtracting $3\sigma_{\text{Cu/Fe}} \approx \SI{19}{\angstrom}$
	\end{tablenotes}
	\end{threeparttable}
\end{table}

Table \ref{table:comparison_layer_parameters} compares the structural parameters of the Si substrate and Cu layer retrieved by a phase reconstruction with the parameters obtained from traditional fitting. 
The Cu layer thicknesses are consistent with the fitted parameter as they differ by a mere $\approx\SI{6}{\angstrom}$. 
It is noted that the reference layer includes the Cu/Fe roughness/interdiffusion transition, which causes the reconstructed Cu layer to be $3\sigma_{\text{Cu/Fe}} \approx \SI{19}{\angstrom}$ thinner. 
%The thickness is calculated by the width of two consecutive jump discontinuities from a step function approximation of the SLD.
The rms roughness at the Si/Cu interface substantially differs from the fitted roughness. 
The deviation can be explained by the fact that the fitting model allowed the resolution to be varied, whereas for reconstruction of the reflection the resolution is not taken into account, see \ref{appendix:resolution}. 

The densities of the Si substrate and the Cu layer are estimated by the average SLD. 
They agree with the fitted densities and are in line with the values reported in the literature.
The noise inherent in the reflection which is reconstructed using only spin-up measurements is preventing a precise inversion of the SLD.

%\hl{A different conclusion can be drawn when reversing the role of the \textit{a-priori} knowledge of the unknown and reference layers: Assuming we know the shape of the Cu layer, it is possible to infer information from the reference layers. For example, assuming that the reference layers have the same constant roughness at the Fe/Air interface we retrieve an inverted scattering potential with a substrate SLD differing by $61\%$ relative to the theoretical Si SLD while a non-constant roughness yields a difference of $10\%$. This shows that the assumption of a constant roughness is not supported by the data. (the plot is not shown though)}

By extrapolating the reflection to higher $q$ a smoothing of the oscillations in the SLD is observed and only small deviations $\leq 4\%$ to the expected model of the SLD inside the Cu layer are identified. To the right of the Si/Cu interface at a depth of $\SI{25}{\angstrom}$ a mismatch of $\approx 13\%$ is visible due to the resolution of the reconstructed reflection. The deviation in the Si SLD of $\approx 10\%$ originates from the mismatch in $\rho_\Delta$, which can be improved by utilizing a more sophisticated model for the reflection extrapolation.

\section{Summary and Conclusion}
We have shown the applicability of the reflection reconstruction with magnetic reference layers to retrieve the reflection of a seed layer for the case of \insitu PNR. Only the top Fe layers are used as a reference layer. On the one hand, splitting the total Fe layer in a remnant and reference part allows the underlying unknown sample to be reconstructed without any further assumptions, but with a reduction of accuracy as only reflectivity measurements of the same spin direction can be used. On the other hand, if one assumes knowledge of the Cu/Fe interface it is possible to use all performed spin-up and spin-down PNR measurements to obtain a more precise reflection.

%For ideal phase-sensitive PNR with magnetic reference layers, we recommend to insert a non-magnetic buffer layer between the sample and the magnetic reference with ideally also known interface parameters to the reference. 
%Moreover, the use of a magnetic reference layer is strongly encouraged since the gain in the SLD contrast for spin-up and down reflectivity increases the accuracy of the reconstructed reflection dramatically.

The parameters extracted of the unknown sample are in general consistent with the values obtained from traditional data fitting. Only the roughness parameters substantially differ from the fitted value which we attribute to the resolution degradation. Nevertheless, we have shown to uniquely extract the desired parameters like roughness, density and thickness of the unknown sample by means of phase-sensitive \insitu PNR.

The presented analysis technique finds its application in the unique determination of depth resolved magnetism such as proximity effects like induced magnetism between heavy-metal and ferromagnetic transition metals (FM) \cite{hase_proximity_2014, mayr_indications_2020}. Additionally, it allows the morphology of thin film system with magnetic tunnel junctions in FM/oxide/FM structures (\eg Fe/MgO/Fe) \cite{ikeda_perpendicular_2010, lambert_quantifying_2013}, or multilayer systems like Fe/Gd or Pt/Co/Ta, which are known to show a formation of skyrmions at room temperature \cite{montoya_tailoring_2017, woo_observation_2016}, to be precisely investigated. It is noted that, although we used ferromagnetic reference layers, the technique of \insitu growth allows non-magnetic reference layers to be applied in order to probe an unknown magnetic part of a sample.

In light of the upgrade of the Selene focusing optics at Amor, PSI and the advent of the new beamline Estia at the European Spallation Source (ESS) a tremendous improvement in the acquired data quality is expected, which will entail a reduction of statistical noise and in turn facilitate even more precisely reconstructed reflections. The limitation of the $q$ range in the reflection will be lessened, as the upgrade allows much higher $q$ values to be assessed at a given data acquisition time, enabling more accurate sample descriptions to be determined. In addition, phase determination in time-resolved polarized neutron reflectometry by \insitu or \textit{in operando} techniques might become viable, as it is already demonstrated for X-ray radiation \cite{kozhevnikov_exact_2008-1, kozhevnikov_exact_2008}.

%\begin{acknowledgments}
\section{Acknowledgments}
This work is based on experiments performed at the Swiss spallation neutron source (SINQ), Paul Scherrer Institute, Villigen, Switzerland. The research is funded by the Deutsche Forschungsgemeinschaft (DFG, German Research Foundation) within the Transregional Collaborative Research Center “From electronic correlations to functionality” - Projekt ID 107745057 - TRR 80.
%\end{acknowledgments}

\renewcommand\thefigure{A.\arabic{figure}}
\setcounter{figure}{0}
\renewcommand{\thetable}{A.\Roman{table}}
\setcounter{table}{0}
\appendix

\section{Influence of Poisson Noise on the Reflection}\label{appendix:fe_down_statistics}

\begin{figure}
	\centering
	\includegraphics[width=\columnwidth]{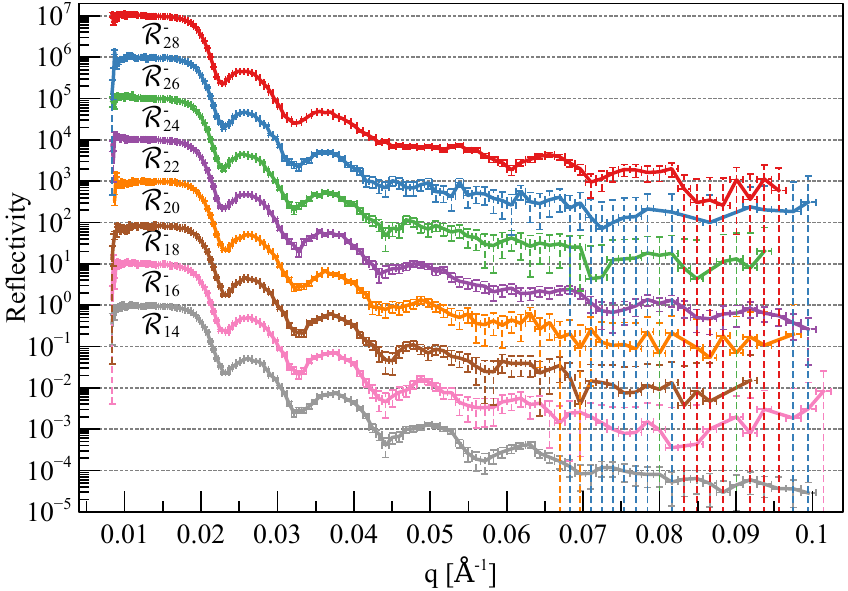}
	\caption{
		Neutron reflectivity data $\Rmeas_n^-$ of the Si/Cu/Fe sample. Each reflectivity is shifted by one order of magnitude for reasons of clarity. The reflectivity $\Rmeas_n^-$ corresponds to the sample with $n$ monolayers of Fe. The sample is magnetized anti-parallel to the neutron spin polarization (spin-down).
	}
	\label{fig:reflectivity_down}
\end{figure}

\begin{figure}
	\centering
	\includegraphics[width=\columnwidth]{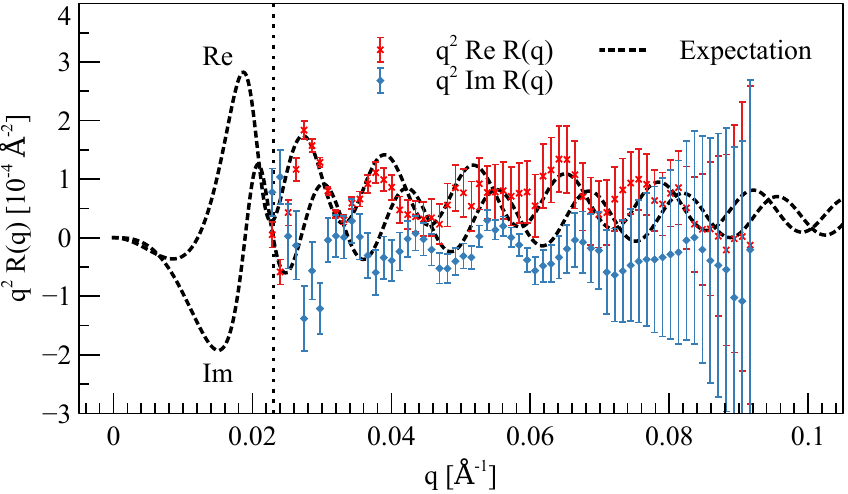}
	\caption{
		Reconstructed reflection multiplied with $q^2$ of the unknown Cu layer, including a remnant Fe layer magnetized anti-parallel to the neutron polarization. The reflection is reconstructed using only the reflectivity data (spin-down) $\Rmeas_{16}^-$ - $\Rmeas_{28}^-$. The expected reflection represents the sample as in Figure \ref{fig:fe_up_remnant_reconstructed_phase} except being magnetized in opposite direction.
	}
	\label{fig:fe_down_remnant_reconstructed_phase}
\end{figure}

To evaluate how Poisson noise in PNR affects the reflection, we reconstructed the reflection $R^-$ using solely spin-down data (Figure \ref{fig:reflectivity_down}), in comparison to $R^+$ (Figure \ref{fig:fe_up_remnant_reconstructed_phase}). The expected reflection (obtained by fitting) and reconstructed reflection is depicted in Figure \ref{fig:fe_down_remnant_reconstructed_phase}. The errors of $R^-$ for $q \geq \SI{0.08}{\angstrom^{-1}}$ encompass nearly the whole range of reflection values and clearly exceed the errors of $R^+$. This observation can be explained by the fact that the spin-down reflectivity decays faster (compared to the spin-up reflectivity) and that the measurement uncertainty originating from the background is more pronounced. %This directly translates to the uncertainty in the reflection $R^-$.
Comparing, however, the spin-up with the spin-down reflection, both coincide with the true reflection equally well, if averaged over the entire measured $q$-range.

The reconstruction for $q$ close to the total reflection edge is highly sensitive to noise in the reflectivity, see Figure \ref{fig:fe_up_remnant_reconstructed_phase} and Figure \ref{fig:fe_down_remnant_reconstructed_phase} in the range $q \in \SIinterval{0.02}{0.03}{\angstrom^{-1}}$. The reflection is transformed by equation \ref{eq:reflectivity_transformation} which has a pole at $|R(q)|^2=1$, \ie at the total reflection edge. Hence, the reconstruction is strongly influenced by noise in the reflectivity data close to the total reflection edge.

\section{Additional Data}\label{appendix:additional_table}
The optimal parameters for the reflection model in Section \ref{sec:reflection_fit} are listed for completeness in Table \ref{table:parameters_fit}. Only the parameters with a given unit are physically interpreted. The parameter $d$ represents the thickness of the Cu layer, $\sigma$ is the rms roughness of the Si/Cu interface, $\rho$ is the SLD of Cu, $\rho_\Delta$ is the difference in the SLD of Cu and Si, and $q_c$ corresponds to the critical wave vector. The values listed in the column ``Expectation'' are obtained from a traditional fit of the reflectivity. 
\begin{table}[h]
	\centering
	\caption{
		Fitted parameters obtained from the reflection model \eqref{eq:reflection_model} with $\chi_{\text{red}}^2 = 1.38$.
	}
	\begin{tabular}{c c c c}
		\toprule
		Symbol 					&	Fit 						& Expectation & Unit \\
		\midrule
		$d$	 				    &	$433 \pm 2$ 				& 433	& $\SI{}{\angstrom}$ \\
		$\sigma$     		  	&	$12.2 \pm 0.4$ 				& 10.7	& $\SI{}{\angstrom}$ \\
		$\rho$					&	$\SI{5.82 \pm 0.04}{}$		& 6.45	& $\SI{e-6}{\angstrom^{-2}}$ \\
		$\rho_\Delta$			&	$\SI{3.94 \pm 0.28}{}$		& 4.38	& $\SI{e-6}{\angstrom^{-2}}$ \\
		$q_c$        			&	$0.019 \pm 0.001$			& 0.018 & $\SI{}{\angstrom^{-1}}$ \\
		\midrule
		$a_0$					&	$\SI{1.5 \pm 0.4e-7}{}$		& - 	& - \\
		$a_1$        			&	$\SI{7.8 \pm 0.2e-6}{}$		& - 	& - \\
		$\phi_0$ 				&	$\SI{0.12 \pm 0.03}{\pi}$	& 0 	& - \\
		$\phi_1$    			&	$\SI{-9.3 \pm 0.2e-2}{}$	& - 	& - \\
		\bottomrule
	\end{tabular}
	\label{table:parameters_fit}
\end{table}

\section{Resolution Effects}\label{appendix:resolution}
\begin{figure}
	\centering
	\includegraphics[width=\columnwidth]{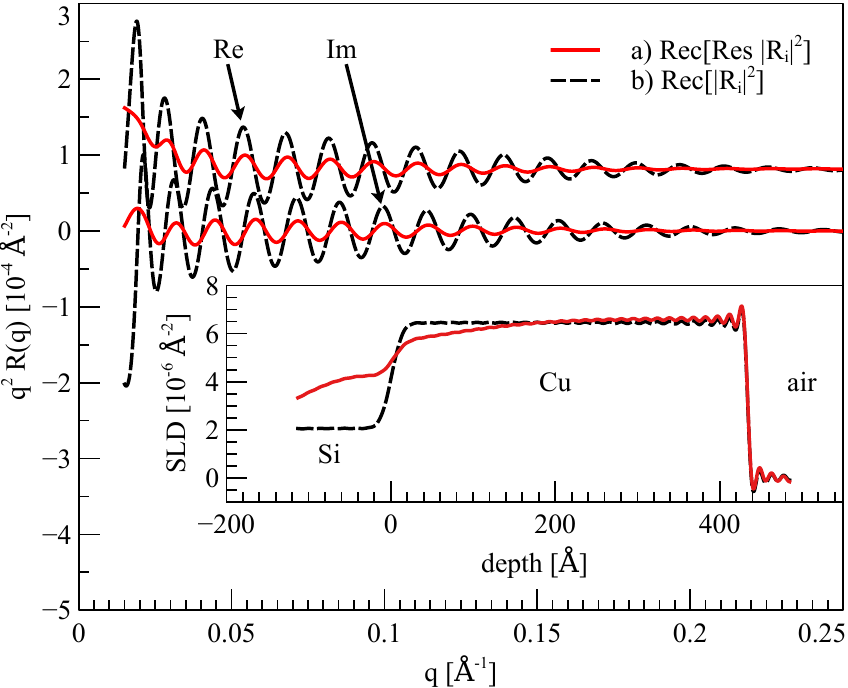}
	\caption{
		Simulation of the influence of resolution in reflectivity measurements onto the reconstructed reflection. a) Shows the reconstructed reflection multiplied with $q^2$ obtained by resolution degraded reflectivity measurements. b) Shows the reconstruction without any degradation. The inset graph depicts the inverted potential of the reflections for $q \in \SIinterval{0}{0.43}{\angstrom^{-1}}$. The true potential is that of a Si substrate with a $\approx\SI{433}{\angstrom}$ thick Cu layer on top.
	}
	\label{fig:resolution_comparison}
\end{figure}

For evaluating the effect of resolution on the reconstructed reflection, we simulated the reflectivity first without including any resolution and then with a constant relative wavelength and constant relative angular resolution. The resolution is given by
\begin{equation*}
\Delta q = q\sqrt{\left(\frac{\Delta \lambda}{\lambda}\right)^2 + \left(\frac{\Delta \theta}{\tan(\theta)}\right)^2}.
\end{equation*}
The resolution of the reflectivity $\operatorname{Res}|R|^2$ is computed by the convolution (denoted by $*$) with a normal distribution $\mathcal{N}_{\Delta q}$ with mean $\mu = 0$ and standard deviation $\sigma = \Delta q(q)$ by
\begin{equation*}
\operatorname{Res}\left[|R|^2\right](q) = \left(|R|^2 * \mathcal{N}_{\Delta q}\right)(q).
\end{equation*}
The reflection is reconstructed as in Section \ref{sec:combination}, however, using the simulated reflectivity data with and without applied resolution. Taking a constant $\Delta \theta_{\text{FWHM}} \equiv \SI{0.3}{\degree}$ and $\Delta \lambda_{\text{FWHM}} \equiv \SI{0.2}{\angstrom}$ the reflection together with the corresponding inverted potential is shown in Figure \ref{fig:resolution_comparison}.

In this simple example we observe a damping ($\approx 1:3$) of the oscillations and a small phase shift ($\Delta \varphi \approx \SI{\pi/13}{\radian}$) in the reconstructed reflection. The resolution degraded potential exhibits a slightly higher ($\approx 2\%$) SLD at the Cu/Fe interface while at the Si/Cu interface the SLD drops as the relative error increases to $10\%$. At the Si substrate the SLD has a mismatch of $> 100\%$, however the thickness of the Cu layer can clearly be identified.

% We conclude that the resolution is not causing any unexpected high frequency components in the reconstructed reflection, but, instead, reduces the amplitude of the reflection.

\begin{figure}
	\centering
	\includegraphics[width=\columnwidth]{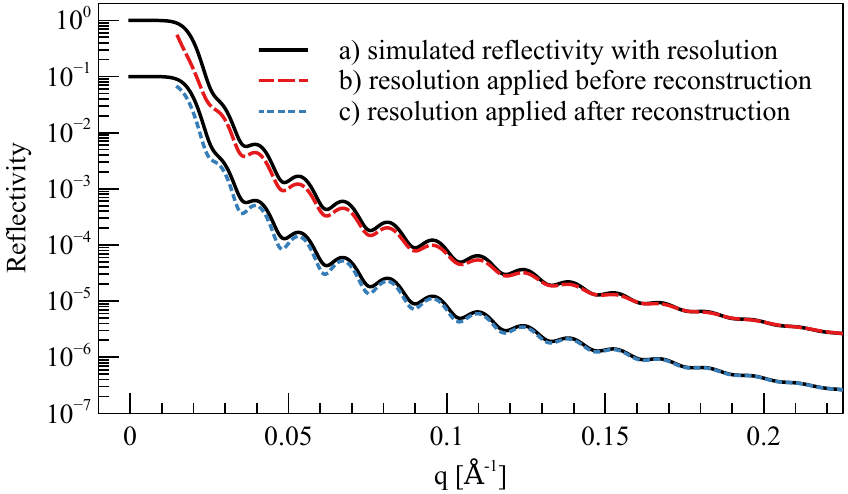}
	\caption{
		Computed reflectivities obtained by a) a common reflectivity simulation b) reconstruction of the reflection with resolution degraded reflectivity data c) reconstruction the reflection with perfect reflectivity data and applying the resolution operator afterwards. The reflectivity is shifted by one order of magnitude. Note that a), b) and c) correspond to $\operatorname{Res} \left|R\right|^2$, $\operatorname{Rec}[\operatorname{Res} (|R_i|^2)]$ and $\operatorname{Res^*}[\operatorname{Rec} (|R_i|^2)]$, respectively.
	}
	\label{fig:resolution_comparison_reflectivity}
\end{figure}

Another interesting observation can be seen in Figure \ref{fig:resolution_comparison_reflectivity}. It shows that the reconstruction operator $\operatorname{Rec}$ and the resolution operator $\operatorname{Res}$ cannot be interchanged, \ie none of the following reflections are equal:
\begin{equation*}
\operatorname{Rec}[\operatorname{Res} (|R_i|^2)], \quad \operatorname{Res^*}[\operatorname{Rec} (|R_i|^2)] , \quad \operatorname{Res} \left|R\right|^2
\end{equation*}
where $\operatorname{Rec}$ denotes the operator which reconstructs the reflection based on $i \in \N, i \geq 3$ reflectivity measurements $|R_i|^2$ and known reference layers, $|R|^2$ denotes the reflectivity of the sample. Recall that $R$ corresponds to the reflection of the unknown sample whereas $R_i$ corresponds to the reflection of the unknown sample with a reference layer $i$. The operator $\operatorname{Res^*}$ denotes the naive definition of a resolution operator by convoluting the real and imaginary part separately. Note that the equation $\operatorname{Rec} (|R_i|^2) = R$ does hold. Especially, it is not possible to reconstruct the true reflection of the unknown sample from resolution degraded reflectivity data.

It is not straightforward and as such not yet clear how to best remove the resolution from the reflection data since a precise description of how the resolution of an instrument changes the reflection of the sample is not known. A first approach is to remove the resolution already from the reflectivity, \eg by deconvolution techniques, and then retrieve the reflection. However, a thorough analysis of this problem is beyond the scope of this work, and will be presented elsewhere.

\section{Error Estimation}\label{appendix:error_estimation}
Since the error propagation and estimation are highly involved, we estimated the errors by numerical Monte-Carlo simulations. The errors of the reflection (for $q$ larger than the minimal wave vector transfer $q_c$) are calculated by the covariance matrix of the minima of $\chi^2$ in equation \eqref{eq:chi_squared}. The errors for the low $q$ region ($q \leq q_c$) are estimated as follows:
From the reflection $R \pm \sigma_R$ one randomly draws $2000$ samples $R'$, from which the reflection $R'(q\leq q_c)$ is retrieved with a fixed-point algorithm \cite{book_retrieval_2020}. The error in $R(q\leq q_c)$ is then set as the standard deviation of $R'$. Likewise, the error in the SLD is the standard deviation of $5000$ inverted, randomly drawn reflections.

It is noted that this error calculation of the SLD might not be precise: The error in $R(q\leq q_c)$ is calculated from $R(q\geq q_c)$ and thus the errors are in general no longer independent for each $q$, contrary to our assumption. 

Note that the error calculations performed in this paper do not include possible errors in the parameters of the reference layers (Table  \ref{table:reference_layer_parameters}).

\bibliographystyle{elsarticle-num-names}
\bibliography{document}

\providecommand{\noopsort}[1]{}\providecommand{\singleletter}[1]{#1}%
\begin{thebibliography}{36}
\expandafter\ifx\csname natexlab\endcsname\relax\def\natexlab#1{#1}\fi
\providecommand{\url}[1]{\texttt{#1}}
\providecommand{\href}[2]{#2}
\providecommand{\path}[1]{#1}
\providecommand{\DOIprefix}{doi:}
\providecommand{\ArXivprefix}{arXiv:}
\providecommand{\URLprefix}{URL: }
\providecommand{\Pubmedprefix}{pmid:}
\providecommand{\doi}[1]{\href{http://dx.doi.org/#1}{\path{#1}}}
\providecommand{\Pubmed}[1]{\href{pmid:#1}{\path{#1}}}
\providecommand{\bibinfo}[2]{#2}
\ifx\xfnm\relax \def\xfnm[#1]{\unskip,\space#1}\fi
%Type = Article
\bibitem[{Fiedeldey et~al.(1992)Fiedeldey, Lipperheide, Leeb, and
  Sofianos}]{fiedeldey_proposal_1992}
\bibinfo{author}{H.~Fiedeldey}, \bibinfo{author}{R.~Lipperheide},
  \bibinfo{author}{H.~Leeb}, \bibinfo{author}{S.~Sofianos},
\newblock \bibinfo{title}{A proposal for the determination of the phases in
  specular neutron reflection},
\newblock \bibinfo{journal}{Physics Letters A} \bibinfo{volume}{170}
  (\bibinfo{year}{1992}) \bibinfo{pages}{347--351}.
  \DOIprefix\doi{10.1016/0375-9601(92)90885-P}.
%Type = Article
\bibitem[{Clinton(1993)}]{clinton_phase_1993}
\bibinfo{author}{W.~L. Clinton},
\newblock \bibinfo{title}{Phase determination in x-ray and neutron reflectivity
  using logarithmic dispersion relations},
\newblock \bibinfo{journal}{Physical Review B} \bibinfo{volume}{48}
  (\bibinfo{year}{1993}) \bibinfo{pages}{1--5}.
  \DOIprefix\doi{10.1103/PhysRevB.48.1}.
%Type = Article
\bibitem[{Sanyal et~al.(1993)Sanyal, Sinha, Gibaud, Huang, Carvalho,
  Rafailovich, Sokolov, Zhao, and Zhao}]{sanyal_fourier_1993}
\bibinfo{author}{M.~K. Sanyal}, \bibinfo{author}{S.~K. Sinha},
  \bibinfo{author}{A.~Gibaud}, \bibinfo{author}{K.~G. Huang},
  \bibinfo{author}{B.~L. Carvalho}, \bibinfo{author}{M.~Rafailovich},
  \bibinfo{author}{J.~Sokolov}, \bibinfo{author}{X.~Zhao},
  \bibinfo{author}{W.~Zhao},
\newblock \bibinfo{title}{Fourier {Reconstruction} of {Density} {Profiles} of
  {Thin} {Films} {Using} {Anomalous} {X}-{Ray} {Reflectivity}},
\newblock \bibinfo{journal}{Europhysics Letters (EPL)} \bibinfo{volume}{21}
  (\bibinfo{year}{1993}) \bibinfo{pages}{691--696}.
  \DOIprefix\doi{10.1209/0295-5075/21/6/010}.
%Type = Article
\bibitem[{Lynn and Seeger(1990)}]{lynn_resonance_1990}
\bibinfo{author}{J.~E. Lynn}, \bibinfo{author}{P.~A. Seeger},
\newblock \bibinfo{title}{Resonance effects in neutron scattering lengths of
  rare-earth nuclides},
\newblock \bibinfo{journal}{Atomic Data and Nuclear Data Tables}
  \bibinfo{volume}{44} (\bibinfo{year}{1990}) \bibinfo{pages}{191--207}.
  \DOIprefix\doi{10.1016/0092-640X(90)90013-A}.
%Type = Article
\bibitem[{Salamatov and Kravtsov(2016)}]{salamatov_use_2016}
\bibinfo{author}{Y.~A. Salamatov}, \bibinfo{author}{E.~A. Kravtsov},
\newblock \bibinfo{title}{Use of gadolinium as a reference layer for neutron
  reflectometry},
\newblock \bibinfo{journal}{Journal of Surface Investigation. X-ray,
  Synchrotron and Neutron Techniques} \bibinfo{volume}{10}
  (\bibinfo{year}{2016}) \bibinfo{pages}{1169--1172}.
  \DOIprefix\doi{10.1134/S1027451016050591}.
%Type = Article
\bibitem[{Nikova et~al.(2019)Nikova, Salamatov, Kravtsov, Makarova, Proglyado,
  Ustinov, Bodnarchuk, and Nagorny}]{nikova_experimental_2019}
\bibinfo{author}{E.~S. Nikova}, \bibinfo{author}{Y.~A. Salamatov},
  \bibinfo{author}{E.~A. Kravtsov}, \bibinfo{author}{M.~V. Makarova},
  \bibinfo{author}{V.~V. Proglyado}, \bibinfo{author}{V.~V. Ustinov},
  \bibinfo{author}{V.~I. Bodnarchuk}, \bibinfo{author}{A.~V. Nagorny},
\newblock \bibinfo{title}{Experimental {Approbation} of {Reference} {Layer}
  {Method} in {Resonant} {Neutron} {Reflectometry}},
\newblock \bibinfo{journal}{Physics of Metals and Metallography}
  \bibinfo{volume}{120} (\bibinfo{year}{2019}) \bibinfo{pages}{838--843}.
  \DOIprefix\doi{10.1134/S0031918X19090102}.
%Type = Article
\bibitem[{Nikova et~al.(2020)Nikova, Salamatov, Kravtsov, Ustinov, Bodnarchuk,
  and Nagorny}]{nikova_development_2020}
\bibinfo{author}{E.~S. Nikova}, \bibinfo{author}{Y.~A. Salamatov},
  \bibinfo{author}{E.~A. Kravtsov}, \bibinfo{author}{V.~V. Ustinov},
  \bibinfo{author}{V.~I. Bodnarchuk}, \bibinfo{author}{A.~V. Nagorny},
\newblock \bibinfo{title}{Development of {Reference} {Layer} {Method} in
  {Resonant} {Neutron} {Reflectometry}},
\newblock \bibinfo{journal}{Journal of Surface Investigation: X-ray,
  Synchrotron and Neutron Techniques} \bibinfo{volume}{14}
  (\bibinfo{year}{2020}) \bibinfo{pages}{S161--S164}.
  \DOIprefix\doi{10.1134/S1027451020070344}.
%Type = Article
\bibitem[{Majkrzak and Berk(1995)}]{majkrzak_exact_1995}
\bibinfo{author}{C.~F. Majkrzak}, \bibinfo{author}{N.~F. Berk},
\newblock \bibinfo{title}{Exact determination of the phase in neutron
  reflectometry},
\newblock \bibinfo{journal}{Physical Review B} \bibinfo{volume}{52}
  (\bibinfo{year}{1995}) \bibinfo{pages}{10827--10830}.
  \DOIprefix\doi{10.1103/PhysRevB.52.10827}.
%Type = Article
\bibitem[{de~Haan et~al.(1995)de~Haan, van Well, Adenwalla, and
  Felcher}]{de_haan_retrieval_1995}
\bibinfo{author}{V.-O. de~Haan}, \bibinfo{author}{A.~A. van Well},
  \bibinfo{author}{S.~Adenwalla}, \bibinfo{author}{G.~P. Felcher},
\newblock \bibinfo{title}{Retrieval of phase information in neutron
  reflectometry},
\newblock \bibinfo{journal}{Physical Review B} \bibinfo{volume}{52}
  (\bibinfo{year}{1995}) \bibinfo{pages}{10831--10833}.
  \DOIprefix\doi{10.1103/PhysRevB.52.10831}.
%Type = Article
\bibitem[{Kirby et~al.(2012)Kirby, Kienzle, Maranville, Berk, Krycka, Heinrich,
  and Majkrzak}]{kirby_phase-sensitive_2012}
\bibinfo{author}{B.~J. Kirby}, \bibinfo{author}{P.~A. Kienzle},
  \bibinfo{author}{B.~B. Maranville}, \bibinfo{author}{N.~F. Berk},
  \bibinfo{author}{J.~Krycka}, \bibinfo{author}{F.~Heinrich},
  \bibinfo{author}{C.~F. Majkrzak},
\newblock \bibinfo{title}{Phase-sensitive specular neutron reflectometry for
  imaging the nanometer scale composition depth profile of thin-film
  materials},
\newblock \bibinfo{journal}{Current Opinion in Colloid \& Interface Science}
  \bibinfo{volume}{17} (\bibinfo{year}{2012}) \bibinfo{pages}{44--53}.
  \DOIprefix\doi{10.1016/j.cocis.2011.11.001}.
%Type = Article
\bibitem[{Majkrzak et~al.(1998)Majkrzak, Berk, Dura, Satija, Karim, Pedulla,
  and Deslattes}]{majkrzak_phase_1998}
\bibinfo{author}{C.~F. Majkrzak}, \bibinfo{author}{N.~F. Berk},
  \bibinfo{author}{J.~A. Dura}, \bibinfo{author}{S.~K. Satija},
  \bibinfo{author}{A.~Karim}, \bibinfo{author}{J.~Pedulla},
  \bibinfo{author}{R.~D. Deslattes},
\newblock \bibinfo{title}{Phase determination and inversion in specular neutron
  reflectometry},
\newblock \bibinfo{journal}{Physica B: Condensed Matter} \bibinfo{volume}{248}
  (\bibinfo{year}{1998}) \bibinfo{pages}{338--342}.
  \DOIprefix\doi{10.1016/S0921-4526(98)00260-9}.
%Type = Article
\bibitem[{Majkrzak et~al.(2009)Majkrzak, Berk, Kienzle, and
  Perez-Salas}]{majkrzak_progress_2009}
\bibinfo{author}{C.~F. Majkrzak}, \bibinfo{author}{N.~F. Berk},
  \bibinfo{author}{P.~Kienzle}, \bibinfo{author}{U.~Perez-Salas},
\newblock \bibinfo{title}{Progress in the {Development} of {Phase}-{Sensitive}
  {Neutron} {Reflectometry} {Methods}},
\newblock \bibinfo{journal}{Langmuir} \bibinfo{volume}{25}
  (\bibinfo{year}{2009}) \bibinfo{pages}{4154--4161}.
  \DOIprefix\doi{10.1021/la802838t}.
%Type = Article
\bibitem[{Majkrzak et~al.(2000)Majkrzak, Berk, Silin, and
  Meuse}]{majkrzak_experimental_2000}
\bibinfo{author}{C.~F. Majkrzak}, \bibinfo{author}{N.~F. Berk},
  \bibinfo{author}{V.~Silin}, \bibinfo{author}{C.~W. Meuse},
\newblock \bibinfo{title}{Experimental demonstration of phase determination in
  neutron reflectometry by variation of the surrounding media},
\newblock \bibinfo{journal}{Physica B: Condensed Matter} \bibinfo{volume}{283}
  (\bibinfo{year}{2000}) \bibinfo{pages}{248--252}.
  \DOIprefix\doi{10.1016/S0921-4526(99)01985-7}.
%Type = Article
\bibitem[{Leeb et~al.(1998)Leeb, Kasper, and
  Lipperheide}]{leeb_determination_1998}
\bibinfo{author}{H.~Leeb}, \bibinfo{author}{J.~Kasper},
  \bibinfo{author}{R.~Lipperheide},
\newblock \bibinfo{title}{Determination of the phase in neutron reflectometry
  by polarization measurements},
\newblock \bibinfo{journal}{Physics Letters A} \bibinfo{volume}{239}
  (\bibinfo{year}{1998}) \bibinfo{pages}{147--152}.
  \DOIprefix\doi{10.1016/S0375-9601(97)00972-9}.
%Type = Article
\bibitem[{Kasper et~al.(1998)Kasper, Leeb, and Lipperheide}]{kasper_phase_1998}
\bibinfo{author}{J.~Kasper}, \bibinfo{author}{H.~Leeb},
  \bibinfo{author}{R.~Lipperheide},
\newblock \bibinfo{title}{Phase determination in spin-polarized neutron
  specular reflection},
\newblock \bibinfo{journal}{Phys. Rev. Lett.} \bibinfo{volume}{80}
  (\bibinfo{year}{1998}) \bibinfo{pages}{2614--2617}.
  \DOIprefix\doi{10.1103/PhysRevLett.80.2614}.
%Type = Article
\bibitem[{Majkrzak et~al.(2003)Majkrzak, Berk, and
  Perez-Salas}]{majkrzak_phase-sensitive_2003}
\bibinfo{author}{C.~F. Majkrzak}, \bibinfo{author}{N.~F. Berk},
  \bibinfo{author}{U.~A. Perez-Salas},
\newblock \bibinfo{title}{Phase-{Sensitive} {Neutron} {Reflectometry}},
\newblock \bibinfo{journal}{Langmuir} \bibinfo{volume}{19}
  (\bibinfo{year}{2003}) \bibinfo{pages}{7796--7810}.
  \DOIprefix\doi{10.1021/la0341254}.
%Type = Article
\bibitem[{Schmehl et~al.(2018)Schmehl, Mairoser, Herrnberger, Stephanos, Meir,
  Förg, Wiedemann, Böni, Mannhart, and Kreuzpaintner}]{schmehl_design_2018}
\bibinfo{author}{A.~Schmehl}, \bibinfo{author}{T.~Mairoser},
  \bibinfo{author}{A.~Herrnberger}, \bibinfo{author}{C.~Stephanos},
  \bibinfo{author}{S.~Meir}, \bibinfo{author}{B.~Förg},
  \bibinfo{author}{B.~Wiedemann}, \bibinfo{author}{P.~Böni},
  \bibinfo{author}{J.~Mannhart}, \bibinfo{author}{W.~Kreuzpaintner},
\newblock \bibinfo{title}{Design and realization of a sputter deposition system
  for the in situ- and in operando-use in polarized neutron reflectometry
  experiments},
\newblock \bibinfo{journal}{Nuclear Instruments and Methods in Physics Research
  Section A: Accelerators, Spectrometers, Detectors and Associated Equipment}
  \bibinfo{volume}{883} (\bibinfo{year}{2018}) \bibinfo{pages}{170--182}.
  \DOIprefix\doi{10.1016/j.nima.2017.11.086}.
%Type = Article
\bibitem[{Ye et~al.(2020)Ye, Book, Mayr, Gabold, Meng, Schäfferer, Need,
  Gilbert, Saerbeck, Stahn, Böni, and Kreuzpaintner}]{ye_design_2020}
\bibinfo{author}{J.~Ye}, \bibinfo{author}{A.~Book}, \bibinfo{author}{S.~Mayr},
  \bibinfo{author}{H.~Gabold}, \bibinfo{author}{F.~Meng},
  \bibinfo{author}{H.~Schäfferer}, \bibinfo{author}{R.~Need},
  \bibinfo{author}{D.~Gilbert}, \bibinfo{author}{T.~Saerbeck},
  \bibinfo{author}{J.~Stahn}, \bibinfo{author}{P.~Böni},
  \bibinfo{author}{W.~Kreuzpaintner},
\newblock \bibinfo{title}{Design and realization of a sputter deposition system
  for the in situ and in operando use in polarized neutron reflectometry
  experiments: {Novel} capabilities},
\newblock \bibinfo{journal}{Nuclear Instruments and Methods in Physics Research
  Section A: Accelerators, Spectrometers, Detectors and Associated Equipment}
  \bibinfo{volume}{964} (\bibinfo{year}{2020}) \bibinfo{pages}{163710}.
  \DOIprefix\doi{10.1016/j.nima.2020.163710}.
%Type = Article
\bibitem[{Kreuzpaintner et~al.(2017)Kreuzpaintner, Wiedemann, Stahn, Moulin,
  Mayr, Mairoser, Schmehl, Herrnberger, Korelis, Haese, Ye, Pomm, Böni, and
  Mannhart}]{kreuzpaintner_situ_2017}
\bibinfo{author}{W.~Kreuzpaintner}, \bibinfo{author}{B.~Wiedemann},
  \bibinfo{author}{J.~Stahn}, \bibinfo{author}{J.-F. Moulin},
  \bibinfo{author}{S.~Mayr}, \bibinfo{author}{T.~Mairoser},
  \bibinfo{author}{A.~Schmehl}, \bibinfo{author}{A.~Herrnberger},
  \bibinfo{author}{P.~Korelis}, \bibinfo{author}{M.~Haese},
  \bibinfo{author}{J.~Ye}, \bibinfo{author}{M.~Pomm},
  \bibinfo{author}{P.~Böni}, \bibinfo{author}{J.~Mannhart},
\newblock \bibinfo{title}{\textit{{In} situ} {Polarized} {Neutron}
  {Reflectometry}: {Epitaxial} {Thin}-{Film} {Growth} of {Fe} on {Cu}(001) by
  dc {Magnetron} {Sputtering}},
\newblock \bibinfo{journal}{Physical Review Applied} \bibinfo{volume}{7}
  (\bibinfo{year}{2017}) \bibinfo{pages}{054004}.
  \DOIprefix\doi{10.1103/PhysRevApplied.7.054004}.
%Type = Article
\bibitem[{Stahn and Glavic(2016)}]{stahn_focusing_2016}
\bibinfo{author}{J.~Stahn}, \bibinfo{author}{A.~Glavic},
\newblock \bibinfo{title}{Focusing neutron reflectometry: {Implementation} and
  experience on the {TOF}-reflectometer {Amor}},
\newblock \bibinfo{journal}{Nuclear Instruments and Methods in Physics Research
  Section A: Accelerators, Spectrometers, Detectors and Associated Equipment}
  \bibinfo{volume}{821} (\bibinfo{year}{2016}) \bibinfo{pages}{44--54}.
  \DOIprefix\doi{10.1016/j.nima.2016.03.007}.
%Type = Article
\bibitem[{Sacks(1993)}]{sacks_reconstruction_1993}
\bibinfo{author}{P.~E. Sacks},
\newblock \bibinfo{title}{Reconstruction of steplike potentials},
\newblock \bibinfo{journal}{Wave Motion} \bibinfo{volume}{18}
  (\bibinfo{year}{1993}) \bibinfo{pages}{21--30}.
  \DOIprefix\doi{10.1016/0165-2125(93)90058-N}.
%Type = Book
\bibitem[{Chadan and Sabatier(1989)}]{chadan_inverse_1989}
\bibinfo{author}{K.~Chadan}, \bibinfo{author}{P.~C. Sabatier},
  \bibinfo{title}{Inverse problems in quantum scattering theory}, Texts and
  monographs in physics, \bibinfo{edition}{2nd ed., rev. and expanded} ed.,
  \bibinfo{publisher}{Springer-Verlag}, \bibinfo{address}{New York},
  \bibinfo{year}{1989}. \DOIprefix\doi{10.1007/978-3-662-12125-2}.
%Type = Misc
\bibitem[{{University of Maryland, DANSE Reflectometry Group}(009 )}]{direfl}
\bibinfo{author}{{University of Maryland, DANSE Reflectometry Group}},
  \bibinfo{title}{{DiRefl} {(Direct Inversion Reflectometry)}},
  \bibinfo{howpublished}{(Version 1.1.2) [Computer Software]},
  \bibinfo{year}{2009-}. \URLprefix
  \url{https://github.com/reflectometry/direfl}.
%Type = Misc
\bibitem[{Book(2019)}]{dinv}
\bibinfo{author}{A.~Book}, \bibinfo{title}{Direct{I}nversion (dinv)},
  \bibinfo{howpublished}{(Version 0.1.0) [Computer Software]},
  \bibinfo{year}{2019}. \URLprefix
  \url{https://github.com/TUM-E21-ThinFilms/DirectInversion}.
%Type = Article
\bibitem[{Book and Kienzle(2020)}]{book_retrieval_2020}
\bibinfo{author}{A.~Book}, \bibinfo{author}{P.~A. Kienzle},
\newblock \bibinfo{title}{Retrieval of the complex reflection coefficient below
  the critical edge for neutron reflectometry},
\newblock \bibinfo{journal}{Physica B: Condensed Matter} \bibinfo{volume}{588}
  (\bibinfo{year}{2020}) \bibinfo{pages}{412181}.
  \DOIprefix\doi{10.1016/j.physb.2020.412181}.
%Type = Article
\bibitem[{Koester et~al.(1991)Koester, Rauch, and
  Seymann}]{koester_neutron_1991}
\bibinfo{author}{L.~Koester}, \bibinfo{author}{H.~Rauch},
  \bibinfo{author}{E.~Seymann},
\newblock \bibinfo{title}{Neutron scattering lengths: {A} survey of
  experimental data and methods},
\newblock \bibinfo{journal}{Atomic Data and Nuclear Data Tables}
  \bibinfo{volume}{49} (\bibinfo{year}{1991}) \bibinfo{pages}{65--120}.
  \DOIprefix\doi{10.1016/0092-640X(91)90012-S}.
%Type = Article
\bibitem[{Berk and Majkrzak(2009)}]{berk_statistical_2009}
\bibinfo{author}{N.~F. Berk}, \bibinfo{author}{C.~F. Majkrzak},
\newblock \bibinfo{title}{Statistical {Analysis} of {Phase}-{Inversion}
  {Neutron} {Specular} {Reflectivity}},
\newblock \bibinfo{journal}{Langmuir} \bibinfo{volume}{25}
  (\bibinfo{year}{2009}) \bibinfo{pages}{4132--4144}.
  \DOIprefix\doi{10.1021/la802779r}.
%Type = Article
\bibitem[{Hewitt and Hewitt(1979)}]{hewitt_gibbs-wilbraham_1979}
\bibinfo{author}{E.~Hewitt}, \bibinfo{author}{R.~E. Hewitt},
\newblock \bibinfo{title}{{The Gibbs-Wilbraham phenomenon: An episode in
  fourier analysis}},
\newblock \bibinfo{journal}{Archive for History of Exact Sciences}
  \bibinfo{volume}{21} (\bibinfo{year}{1979}) \bibinfo{pages}{129--160}.
  \DOIprefix\doi{10.1007/BF00330404}.
%Type = Article
\bibitem[{Hase et~al.(2014)Hase, Brewer, Arnalds, Ahlberg, Kapaklis, Björck,
  Bouchenoire, Thompson, Haskel, Choi, Lang, Sánchez-Hanke, and
  Hjörvarsson}]{hase_proximity_2014}
\bibinfo{author}{T.~P.~A. Hase}, \bibinfo{author}{M.~S. Brewer},
  \bibinfo{author}{U.~B. Arnalds}, \bibinfo{author}{M.~Ahlberg},
  \bibinfo{author}{V.~Kapaklis}, \bibinfo{author}{M.~Björck},
  \bibinfo{author}{L.~Bouchenoire}, \bibinfo{author}{P.~Thompson},
  \bibinfo{author}{D.~Haskel}, \bibinfo{author}{Y.~Choi},
  \bibinfo{author}{J.~Lang}, \bibinfo{author}{C.~Sánchez-Hanke},
  \bibinfo{author}{B.~Hjörvarsson},
\newblock \bibinfo{title}{Proximity effects on dimensionality and magnetic
  ordering in pd/fe/pd trialyers},
\newblock \bibinfo{journal}{Physical Review B} \bibinfo{volume}{90}
  (\bibinfo{year}{2014}) \bibinfo{pages}{104403}.
  \DOIprefix\doi{10.1103/PhysRevB.90.104403}.
%Type = Article
\bibitem[{Mayr et~al.(2020)Mayr, Ye, Stahn, Knoblich, Klein, Gilbert, Albrecht,
  Paul, Böni, and Kreuzpaintner}]{mayr_indications_2020}
\bibinfo{author}{S.~Mayr}, \bibinfo{author}{J.~Ye}, \bibinfo{author}{J.~Stahn},
  \bibinfo{author}{B.~Knoblich}, \bibinfo{author}{O.~Klein},
  \bibinfo{author}{D.~A. Gilbert}, \bibinfo{author}{M.~Albrecht},
  \bibinfo{author}{A.~Paul}, \bibinfo{author}{P.~Böni},
  \bibinfo{author}{W.~Kreuzpaintner},
\newblock \bibinfo{title}{Indications for dzyaloshinskii-moriya interaction at
  the pd/fe interface studied by in situ polarized neutron reflectometry},
\newblock \bibinfo{journal}{Physical Review B} \bibinfo{volume}{101}
  (\bibinfo{year}{2020}) \bibinfo{pages}{024404}.
  \DOIprefix\doi{10.1103/PhysRevB.101.024404}.
%Type = Article
\bibitem[{Ikeda et~al.(2010)Ikeda, Miura, Yamamoto, Mizunuma, Gan, Endo, Kanai,
  Hayakawa, Matsukura, and Ohno}]{ikeda_perpendicular_2010}
\bibinfo{author}{S.~Ikeda}, \bibinfo{author}{K.~Miura},
  \bibinfo{author}{H.~Yamamoto}, \bibinfo{author}{K.~Mizunuma},
  \bibinfo{author}{H.~D. Gan}, \bibinfo{author}{M.~Endo},
  \bibinfo{author}{S.~Kanai}, \bibinfo{author}{J.~Hayakawa},
  \bibinfo{author}{F.~Matsukura}, \bibinfo{author}{H.~Ohno},
\newblock \bibinfo{title}{A perpendicular-anisotropy {CoFeB}–{MgO} magnetic
  tunnel junction},
\newblock \bibinfo{journal}{Nature Materials} \bibinfo{volume}{9}
  (\bibinfo{year}{2010}) \bibinfo{pages}{721--724}.
  \DOIprefix\doi{10.1038/nmat2804}.
%Type = Article
\bibitem[{Lambert et~al.(2013)Lambert, Rajanikanth, Hauet, Mangin, Fullerton,
  and Andrieu}]{lambert_quantifying_2013}
\bibinfo{author}{C.-H. Lambert}, \bibinfo{author}{A.~Rajanikanth},
  \bibinfo{author}{T.~Hauet}, \bibinfo{author}{S.~Mangin},
  \bibinfo{author}{E.~E. Fullerton}, \bibinfo{author}{S.~Andrieu},
\newblock \bibinfo{title}{Quantifying perpendicular magnetic anisotropy at the
  fe-{MgO}(001) interface},
\newblock \bibinfo{journal}{Applied Physics Letters} \bibinfo{volume}{102}
  (\bibinfo{year}{2013}) \bibinfo{pages}{122410}.
  \DOIprefix\doi{10.1063/1.4798291}.
%Type = Article
\bibitem[{Montoya et~al.(2017)Montoya, Couture, Chess, Lee, Kent, Henze, Sinha,
  Im, Kevan, Fischer, {McMorran}, Lomakin, Roy, and
  Fullerton}]{montoya_tailoring_2017}
\bibinfo{author}{S.~A. Montoya}, \bibinfo{author}{S.~Couture},
  \bibinfo{author}{J.~J. Chess}, \bibinfo{author}{J.~C.~T. Lee},
  \bibinfo{author}{N.~Kent}, \bibinfo{author}{D.~Henze}, \bibinfo{author}{S.~K.
  Sinha}, \bibinfo{author}{M.-Y. Im}, \bibinfo{author}{S.~D. Kevan},
  \bibinfo{author}{P.~Fischer}, \bibinfo{author}{B.~J. {McMorran}},
  \bibinfo{author}{V.~Lomakin}, \bibinfo{author}{S.~Roy},
  \bibinfo{author}{E.~E. Fullerton},
\newblock \bibinfo{title}{Tailoring magnetic energies to form dipole skyrmions
  and skyrmion lattices},
\newblock \bibinfo{journal}{Physical Review B} \bibinfo{volume}{95}
  (\bibinfo{year}{2017}) \bibinfo{pages}{024415}.
  \DOIprefix\doi{10.1103/PhysRevB.95.024415}.
%Type = Article
\bibitem[{Woo et~al.(2016)Woo, Litzius, Krüger, Im, Caretta, Richter, Mann,
  Krone, Reeve, Weigand, Agrawal, Lemesh, Mawass, Fischer, Kläui, and
  Beach}]{woo_observation_2016}
\bibinfo{author}{S.~Woo}, \bibinfo{author}{K.~Litzius},
  \bibinfo{author}{B.~Krüger}, \bibinfo{author}{M.-Y. Im},
  \bibinfo{author}{L.~Caretta}, \bibinfo{author}{K.~Richter},
  \bibinfo{author}{M.~Mann}, \bibinfo{author}{A.~Krone}, \bibinfo{author}{R.~M.
  Reeve}, \bibinfo{author}{M.~Weigand}, \bibinfo{author}{P.~Agrawal},
  \bibinfo{author}{I.~Lemesh}, \bibinfo{author}{M.-A. Mawass},
  \bibinfo{author}{P.~Fischer}, \bibinfo{author}{M.~Kläui},
  \bibinfo{author}{G.~S.~D. Beach},
\newblock \bibinfo{title}{Observation of room-temperature magnetic skyrmions
  and their current-driven dynamics in ultrathin metallic ferromagnets},
\newblock \bibinfo{journal}{Nature Materials} \bibinfo{volume}{15}
  (\bibinfo{year}{2016}) \bibinfo{pages}{501--506}.
  \DOIprefix\doi{10.1038/nmat4593}.
%Type = Article
\bibitem[{Kozhevnikov et~al.(2008{\natexlab{a}})Kozhevnikov, Peverini, and
  Ziegler}]{kozhevnikov_exact_2008-1}
\bibinfo{author}{I.~Kozhevnikov}, \bibinfo{author}{L.~Peverini},
  \bibinfo{author}{E.~Ziegler},
\newblock \bibinfo{title}{Exact determination of the phase in timeresolved
  x-ray reflectometry},
\newblock \bibinfo{journal}{Optics Express} \bibinfo{volume}{16}
  (\bibinfo{year}{2008}{\natexlab{a}}) \bibinfo{pages}{144}.
  \DOIprefix\doi{10.1364/OE.16.000144}.
%Type = Article
\bibitem[{Kozhevnikov et~al.(2008{\natexlab{b}})Kozhevnikov, Peverini, and
  Ziegler}]{kozhevnikov_exact_2008}
\bibinfo{author}{I.~Kozhevnikov}, \bibinfo{author}{L.~Peverini},
  \bibinfo{author}{E.~Ziegler},
\newblock \bibinfo{title}{Exact solution of the phase problem in in situ x-ray
  reflectometry of a growing layered film},
\newblock \bibinfo{journal}{Journal of Applied Physics} \bibinfo{volume}{104}
  (\bibinfo{year}{2008}{\natexlab{b}}) \bibinfo{pages}{054914}.
  \DOIprefix\doi{10.1063/1.2968218}.

\end{thebibliography}

\end{document}